# Results from Monitoring the Broad-Line Radio Galaxy 3C 390.3

K. M. Leighly[1], M. Dietrich[2], E. Waltman[3], R. Edelson[4], I. George[5], M. Malkan[6], M. Matsuoka[1], R. Mushotzky[5], and B. M. Peterson[7]

[1] Cosmic Radiation Laboratory, RIKEN, Hirosawa 2-1, Wako-shi, Saitama 351 Japan
[2] Landessternwarte Heidelberg, Königstuhl, D-69117 Heidelberg, Germany
[3] Naval Research Laboratory, Code 7214, Washington, DC 20375-5351, USA
[4] University of Iowa, Dept. of Physics and Astronomy, 203 Van Allen Hall, Iowa City, Iowa, 52242, USA
[5] NASA/Goddard Space Flight Center, Code 660, Greenbelt, MD, 20771, USA
[6] Univ. of California, Dept. of Astronomy, Los Angeles, CA 90024-1562, USA
[7] Ohio State Univ., Dept. of Astronomy, 174 W. 18th Ave., Columbus, OH 43210-1106, USA

**Abstract.** During 1995, the broad-line radio galaxy 3C 390.3 is the subject of a multi-wavelength monitoring campaign comprised of ROSAT HRI, IUE, and ground based optical, infrared and radio observations. We report preliminary results from the monitoring campaign focusing on the X-ray observations. Snapshot ROSAT observations being made every three days show large amplitude variability. The light curve is dominated by a flare near JD 2449800 characterized by a doubling time scale of 9 days and a general increase in flux after the flare. The optical R and I band light curves show a general increase in flux. The ASCA spectra obtained before and after the flare can be described by an absorbed power law. Spectral variability between the two observations is characterized by an increase in power law index by $\Delta\Gamma \sim 0.08$ at higher flux.

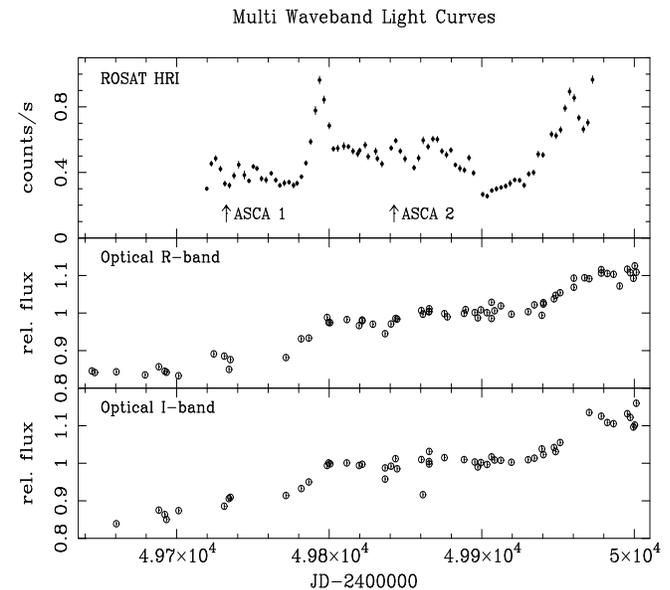

**Fig. 1.** Soft X-ray and optical R and I-band light curves from 1995 3C 390.3 monitoring campaign. Times of two 20ks ASCA observations are also marked.

## 1. Introduction

3C 390.3 is a luminous ($L_{X(2-10)} \sim 6 \times 10^{44}\,\mathrm{ergs\,s^{-1}}$) nearby ($z = 0.057$) broad-line radio galaxy located in the north ecliptic cap, notable for its variable broad, double-peaked Balmer lines possibly produced in an accretion disk (e.g. Perez et al. 1988). The continuum spectrum from 3C 390.3 is complex, with evidence for a thermal component [optical variable blue excess (Shafer, Ward & Barr 1985); soft X-ray excess (Urry et al. 1989; Ghosh et al. 1991, Walter et al. 1994); weak reflection component (Nandra & Pounds 1994)]. A resolved iron fluorescence line was found in the 1993 ASCA spectra which was consistent with production in a Keplerian accretion disk about $250 R_g$ from the center (Eracleous et al. 1995). 3C 390.3 is variable in all wave bands on time scales from days to years, including the optical (Shafer, Ward & Barr 1985) and UV (Clavel & Wamsteker 1987). Hard X-ray variability by a factor of two was reported over a period of 6 weeks (Mushotzky et al. 1977). Five EXOSAT observations show variability by a factor of > 5 in the ME and $\sim 10$ in the LE over the life time of that satellite. A 35% decrease in flux and spectral variability was observed between two Ginga observations 3 years apart (Inda et al. 1994).

## 2. X-ray and Optical Light curves

Figure 1 shows the ROSAT HRI and optical R and I band light curves obtained using the Heidelberg 0.7m telescope (Dietrich et al. in prep.). Radio observations at 2 and 8 GHz made at the NRL-Green Bank Interferometer showed only marginal variability. Each HRI observation is $\sim 1.5$ ks, error bars show $1\sigma$ confidence and the signal-to-noise ranges from 17 to 42. We expect to add eight more points so the ROSAT light curve is nearly complete.

The most remarkable feature is the flare which peaks at TJD 9794 (March 17, 1995). This feature is smooth and it is likely that at least in this region the light curve is resolved. A factor of 3 increase in flux is observed in $\sim 12$ days and a factor of 2 decrease in flux is observed in $\sim 9$ days. The peak flux of 1 HRI count/s corresponds to 0.1–2 keV flux of $3.2 \times 10^{-11}\,\mathrm{ergs\,cm^{-2}\,s^{-1}}$, assuming $\Gamma = 1.78$ and $N_H = 1.3 \times 10^{21}\,\mathrm{cm^{-2}}$ (see below). This corresponds to an intrinsic soft X-ray luminosity at the redshift $z = 0.057$ of $1.1 \times 10^{45}\,\mathrm{ergs\,s^{-1}}$.

The character of the variability observed in the R and I bands is related to but different than that of the soft X-rays. At first glance it may seem that the optical variability is just a smeared version of the faster X-ray variations. However, the X-rays decrease to the pre-flare level by JD 2449900 and remain in a low state for longer than the optical variability time scale ($< 50$ days). It is also possible that the large amplitude X-ray flares are due to a different mechanism than the lower level emission and are simply superimposed on the light curve. However, it can be shown that just before and after the flares the variability amplitude is reduced, implying a causal relationship between the lower level emission and the flares.

## 3. ASCA Observations

We observed 3C 390.3 using ASCA for 20ks twice and the times of the observations are marked on Figure 1. In both cases, the spectra from all four detectors was adequately modeled by an absorbed power law. The absorptions obtained (Table 1) are substantially higher than the Galactic value of $4.2 \times 10^{20}\,\mathrm{cm^{-2}}$ and also somewhat higher than that found by Eracleous et al. (1995) in the 50ks 1993 ASCA observation ($9.7 \pm 1.4 \times 10^{20}\,\mathrm{cm^{-2}}$). However, the larger absorption may be an artifact from changes in the instrument response. A resolved iron line was found in the previous ASCA observation (Eracleous et al. 1995) with $E = 6.34$ keV, $\sigma = 0.15$ keV, $F = 4.0^{+2.3}_{-2.0} \times 10^{-5}\,\mathrm{photons\,cm^{-2}\,s^{-1}}$, and equivalent width $190^{+110}_{-100}$ eV. An iron line is not apparent in power law fit residuals of either 1995 observation, probably because of the poorer statistics due to shorter exposures. If a line is included in the model with parameters fixed at the 1993 values, the F test shows that the iron line is detected at $> 97.5$ and $> 90$% confidence in the first and second observations respectively, and the fit parameters (Table 1) are consistent with those reported by Eracleous et al. (1995).

Between the first and second ASCA observations, the 0.4–10 keV flux increases by a factor of 1.52, while the corresponding HRI observations show a flux change by a factor of 1.77 in the 0.1–2.0 keV band. This suggests that the spectrum varied. Figure 2 shows the $\Gamma$ vs. $N_H$ $\chi^2$

**Table 1.** ASCA Spectral Fitting Results

| Parameter | Obs. 1 | Obs. 2 |
|---|---|---|
| $N_H(\times 10^{21}\,\mathrm{cm^{-2}})$ | $1.4 \pm 0.18$ | $1.2 \pm 0.14$ |
| Photon Index | $1.72 \pm 0.04$ | $1.81 \pm 0.03$ |
| Line flux($\times 10^{-5}\,\mathrm{photons\,cm^{-2}\,s^{-1}}$) | $1.95 \pm 1.7$ | $1.6 < 3.8$ |
| Line Eq. Width (eV) | $99 \pm 89$ | $47 < 113$ |
| $\chi^2$/d.o.f. | 843/825 | 1120/1144 |
| Flux$_{0.4-10\mathrm{keV}}$(ergs cm$^{-2}$ s$^{-1}$) | $2.7 \times 10^{-11}$ | $4.1 \times 10^{-11}$ |
| Intrinsic Luminosity$_{2-10\mathrm{keV}}$(ergs s$^{-1}$) | $2.7 \times 10^{44}$ | $4.0 \times 10^{44}$ |

Notes: Quoted errors are 90% for two parameters of interest ($\Delta \chi^2 = 4.61$). Line energy and width are fixed at 6.34 and 0.15 keV respectively (Eracleous et al. 1995).

contours for both observations. The spectral variability is consistent with an index change at the 90% confidence level. The ASCA spectral parameters projected into the HRI band underpredict the HRI count rates by a factor of $\sim 1.3$, but this may be due to calibration uncertainty.

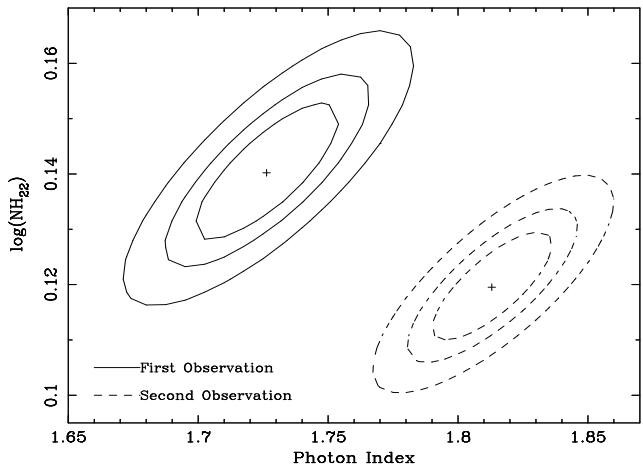

**Fig. 2.** 99%, 90% and 68% $\chi^2$ contours of photon index versus absorption column for the first and second ASCA observations indicate photon index variability at the 90% confidence level.